\documentclass[iop]{emulateapj}

\usepackage{color}
\usepackage[section]{placeins}
\usepackage{graphicx}
\usepackage{epstopdf}
\usepackage{amsmath}
\usepackage{enumitem}

\begin{document}

\title{Multiple measurements of gravitational waves acting as standard probes: model-independent constraints on the cosmic curvature
with DECIGO}

\author{Yilong Zhang\altaffilmark{1}, Shuo Cao\altaffilmark{1$\ast$}, Xiaolin Liu\altaffilmark{1}, Tonghua Liu\altaffilmark{2}, Yuting Liu\altaffilmark{1}, Chenfa Zheng\altaffilmark{1}}

\affiliation{1. Department of Astronomy, Beijing Normal University, Beijing 100875, China; \email{caoshuo@bnu.edu.cn} \\
2. School of Physics and Optoelectronic, Yangtze University, Jingzhou 434023, China}

\begin{abstract}
Although the spatial curvature has been precisely determined via the cosmic microwave background (CMB) observation by Planck satellite, it still suffers from the well-known cosmic curvature tension. As a standard siren, gravitational waves (GWs) from binary neutron star mergers provide a direct way to measure the luminosity distance. In addition, the accelerating expansion of the universe may cause an additional phase shift in the gravitational waveform, which allows us to measure the acceleration parameter. This measurement provides an important opportunity to determine the curvature parameter $\Omega_k$ in the GW domain based on the combination of two different observables for the same objects at high redshifts. In this study, we investigate how such an idea could be implemented with future generation of space-based DECi-hertz Interferometer Gravitational-wave Observatory (DECIGO) in the framework of two model-independent methods. Our results show that DECIGO could provide a reliable and stringent constraint on the cosmic curvature at a precision of $\Delta\Omega_k$=0.12, which is comparable to existing results based on different electromagnetic data. Our constraints are more stringent than the traditional electromagnetic method from the Pantheon SNe Ia sample, which shows no evidence for the deviation from the flat universe at $z\sim 2.3$. More importantly, with our model-independent method, such a second-generation space-based GW detector would also be able to explore the possible evolution $\Omega_k$ with redshifts, through direct measurements of cosmic curvature at different redshifts ($z\sim 5$). Such a model-independent $\Omega_k$ reconstruction to the distance past can become a milestone in gravitational-wave cosmology.

\end{abstract}


\section{Introduction}

The spatial curvature parameter $\Omega_k$ is an important quantity, that is related to many fundamental issues in modern cosmology, such as the structure and evolution of the universe \citep{2006PhRvD..73h3526I,2015PhRvD..91h3536Z,Cao2019a,Qi2019a}. Specifically, the study of the cosmic curvature can effectively test the fundamental assumption of modern cosmology that the universe is homogeneous and isotropic and is well described by the Friedmann-Lemaître-Robertson-Walker (FLRW) metric.
The most popular theory of the very early universe proposes that our universe once went through an exponential phase of expansion, which indicates that the radius of curvature of the universe should be very large and the cosmic curvature should be closed to zero \citep{Weinberg2013}.
Current cosmological observations, particularly the latest \textit{Planck2018} results, which combined the cosmic microwave background (CMB) and baryon acoustic oscillation (BAO) measurements, strongly favor a flat universe: $\Omega_k=0.0007\pm0.0019$ \citep{Planck Collaboration2018} (TT,TE,EE+lowE+lensing+BAO). However, the combination of the \textit{Planck2018} TT,TE,EE+lowE power spectra data alone marginally favors a mildly closed universe: $\Omega_k= -0.044^{+0.018}_{-0.015}$ \citep{Planck Collaboration2018,Valentino2019}.
In addition, such a stringent constraint on the cosmic curvature strongly relies on the assumption of a specific cosmological model (the cosmological constant plus cold dark matter model, i.e., the $\Lambda$CDM model). However, recent analysis indicated that the flat-universe assumption may lead to an incorrect reconstruction of the dark energy equation of state and cause tension between the $\Lambda$CDM and dynamical dark-energy model \citep{Ichikawa2006,Clarkson2007,Gong2007,Virey2008,Li2018,Cao2019b}. The strong degeneracy between the cosmic curvature $\Omega_k$ and the dark energy equation of state $w$ makes it difficult to constrain these two parameters simultaneously in a non-flat $w$CDM model. Therefore, it would be better to measure spatial curvature in geometric and model-independent ways.

The distance sum rule has been proposed as a model-independent method to constrain the curvature of the universe, which is generally implemented using strong lensing observations \citep{Cao12a,Cao12b,Cao15,Ma19} with other distance measurements, such as type Ia supernovae (SNe Ia) \citep{Rasanen2015,Denissenya2018} or quasar (QSO) \citep{Qi2019b,Zhou2020, Lian2021}.
Another model-independent way to constrain the cosmic curvature is comparing the theoretical comoving distance, which is inferred from the observational Hubble parameter data (OHD), with the observed luminosity distances $D_L(z)$ (or the angular diameter distance $D_A(z)$) \citep{Clarkson2008,Shafieloo2010,Mortsell2011,Sapone2014,Cai2016}.
This test has been fully implemented with various observational data.
For example, \citet{Wang2020a} recently used the latest Pantheon SNe Ia data combined with cosmic chronometers (CC) to constrain the cosmic curvature;
a well-measured quasar sample \citep{Risaliti2015,Lusso2016,Risaliti2017,Lusso2017} also shows a great potential for probing the cosmic curvature \citep{Wei2020};
\citet{Cao2019b} proposed an improved model-independent test of cosmic curvature with ultra-compact structures in radio quasars as standard rulers, which shows which shows no evidence for the deviation from the flat universe;
and \citet{Takada2015} proposed that the combination of radial and angular diameter distances from future BAO experiments can be used for studying the curvature parameter.
One can also use an analytical equation to reconstruct $\Omega_k$ at different redshifts \citep{Clarkson2007}, which also requires the observational data of the Hubble parameter $H(z)$ and luminosity distance $D_L(z)$, plus the first derivative of the latter. Such a method helps us study the evolution of the spacial curvature and provides a direct geometric way to test the assumption of cosmic homogeneity and isotropy. $\Omega_k(z)$ reconstructed in this way using current observational data, the combination of cosmic chronometer data with 1598 quasars\citep{Risaliti2018} and the Pantheon catalogue of 1048 SNe Ia\citep{Scolnic2018}, shows good agreement with a flat universe at different redshifts \citep{Liu2020}.

On the other hand, gravitational-wave (GW) observations soon caught people's attention with the discovery of the first GW event, GW150914 \citep{Abbott16}. As standard sirens, GW signals from inspiraling and merging compact binaries encode distance information \citep{Schutz1986}, provide an absolute measurement of the luminosity distance. Only if these binary mergers are accompanied by short-hard $\gamma$-ray bursts (shGRB), can they be observed through both electromagnetic (EM) and GW.
The joint detection of GW170817 \citep{Abbott17} has detected its electromagnetic counterparts from the merger of binary neutron stars (NSs). Knowing the redshifts of the sources, these GW signals can be used for cosmology. Also, the accelerating expansion of the universe would cause an additional phase shift in the gravitational waveform, which allows us to measure the cosmic acceleration (or redshift-drift) directly \citep{Cutler2009, Nishizawa2011}. Thus, it is possible for us to break the degeneracy and measure cosmic curvature independently in GW domain.

However, the measurement of the cosmic acceleration requires a high-precision GW detection, particularly at lower frequencies when the binary remains in its inspiraling phase. For gravitational-wave signals from a neutron star binary at a redshift of 1, the universe expansion acceleration would cause a phase delay of only 1 sec during the 10-year probe \citep{Seto2001, Nishizawa2012}. As a future space-borne GW detector, DECIGO \citep{Kawamura2011,Seto2001} is designed to improve the detection sensitivity of GW at lower frequencies, with its most sensitive frequencies between 0.1 and 10 Hz. DECIGO will have four clusters of spacecraft, and each cluster consists of three spacecraft with three Fabry-Perot Michelson interferometers, whose arm length is 1000 km, improving the determination of their position in the sky. The expected sensitivity of DECIGO is $10^{-25}$ Hz$^{-1/2}$ for two clusters at the same position for three years of mission, which enables the early detection of inspiraling sources. Therefore, DECIGO would create an unprecedented opportunity to precisely measure the cosmic acceleration from GW signals and make GW a more precise standard siren.

Due to the lack of GW events with observed EM counterparts, we simulated 10,000 GW events from DECIGO within the redshift ranges of $0\sim5$. We applied two model-independent methods to measure the cosmic curvature: numerical constraint and reconstruction. For comparison, we also use Pantheon SNe Ia and observational Hubble parameter data (OHD) to estimate $\Omega_k$. The remainder of this paper is organized as follows. In Section 2, we introduce the simulated GW data and methodology used in this study. In Section 3, we present the results of our study and provide some discussion by comparison. Finally, the general conclusions are summarized in Section 4.

\section{Methodology}

Assuming that the universe is homogeneous and isotropic on large scales, it can be described by the FLRW metric:
\begin{equation}
ds^2 = -dt^2 + a^2(t)[\frac{dr^2}{1-Kr^2}+r^2(d\theta^2+sin^2\theta
d\phi^2)],
\end{equation}
where $t$ is the cosmic time and ($r$,$\theta$,$\phi$) are the comoving spatial coordinates. The scale factor $a(t)$ is the only gravitational degree of freedom and its evolution is determined by matter and energy of the universe. The dimensionless curvature $K=-1,0,+1$ corresponds to open, flat and closed universes respectively.
With such a metric, the luminosity distance $D_L(z)$ can be expressed as
\begin{equation}\label{eq2}
D_L(z) = \left\lbrace \begin{array}{lll}
\frac{c(1+z)}{H_0\sqrt{|\Omega_{\rm k}|}}\sinh\left[\sqrt{|\Omega_{\rm k}|}\int_{0}^{z}\frac{dz'}{E(z')}\right]~~{\rm for}~~\Omega_{k}>0,\\
\frac{c(1+z)}{H_0}\int_{0}^{z}\frac{dz'}{E(z')}~~~~~~~~~~~~~~~~~~~~~~~{\rm for}~~\Omega_{K}=0, \\
\frac{c(1+z)}{H_0\sqrt{|\Omega_{\rm k}|}}\sin\left[\sqrt{|\Omega_{\rm k}|}\int_{0}^{z}\frac{dz'}{E(z')}\right]~~~~{\rm for}~~\Omega_{k}<0.\\
\end{array} \right.
\end{equation}
The dimensionless Hubble parameter $E(z)$ is defined as $H(z)/H_0$, where $H(z)$ is the universe expansion rate and $H_0$ is the Hubble constant. The curvature parameter $\Omega_k$ is related to $K$ as
$\Omega_k=-c^2 K/(a_0H_0)^2$, where $c$ is the speed of light.

\subsection{Measuring the distance and the acceleration parameter with GWs}

As a standard siren, GW signal allows us to measure the luminosity distance directly without relying on the cosmic distance ladder. Concurrently, if the expansion of the universe is accelerating, we might find an additional phase shift in the gravitational waveform, from which we can measure the acceleration parameter $X$, and thus obtain the Hubble parameter. Therefore, both the luminosity distance and the Hubble parameter can be obtained from a single GW signal. To apply this $D_L-X$ relation to cosmological studies, we still need information about the redshifts, which requires the GW sources to be neutron-star binaries (NS-NS) or black hole-neutron star binaries (BH-NS), the origins of kilonovae or shGRBs. As long as the corresponding EM counterparts are observed from GW events, we can obtain their redshifts and apply them to cosmology (discussed in Section 2.2).

In this study, we focus on the GW signals from NS-NS binary systems with component masses $m_1$ and $m_2$. One can define the chrip mass $M_c=M\eta^{3/5} $, and the redshifted chirp mass $\mathcal{M}_z\equiv M (1+z_c) \eta^{3/5}$, where $M=m_1+m_2$ is the total mass of the binary system, $\eta= m_1m_2/M^2$ is the symmetric mass ratio, and $z_c$ is the source redshift at coalescence.

We first derive the correction to the GW phase due to the accelerating expansion of the universe. The observed gravitational waveform can be represented by $h(\Delta t)$, where $\Delta t \equiv t_c - t$ denotes the time to coalescence measured in the observer frame, with $t_c$ representing the coalescence time. The Fourier component of this waveform can be written as
\begin{equation}\label{eq5}
\begin{aligned}
\tilde{h}(f) &=\int_{-\infty}^{\infty} e^{2 \pi i f t} h(\Delta t) d t.
\end{aligned}
\end{equation}
Then, we define $\Delta t_e$ as the time to coalescence measured in the source frame, and $\Delta T \equiv (1+z_c) \Delta t_e$ as the redshifted coalescence time. The coalescence times in these two different frames have a relation \footnote{Such a relation is only an approximation, without considering the contribution from high-order terms.}:
\begin{equation}\label{eq6}
\Delta t=\Delta T+X\left(z_{c}\right) \Delta T^{2},
\end{equation}
which is measured by the acceleration parameter $X(z)$ \citep{Seto2001,Takahashi2005} and the correction term due to cosmic acceleration.
The acceleration parameter $X(z)$ is defined as \footnote{Note that $X(z)$ is related to the redshift drift $\Delta_t z$ as $\Delta_t z = H_0 \Delta t_o \left( 1+z-\frac{H(z)}{H_0} \right)$ in FLRW spacetime.}
\begin{equation}
X(z) \equiv \frac{H_0}{2} \left(1-\frac{H(z)}{(1+z) H_0} \right).
\end{equation}
By substituting Eq.~(4) into Eq.~(3) and applying the stationary phase approximation \citep{Cutler1994}, we can transform the Fourier component of the waveform into the frequency domain
\begin{equation}
\tilde{h}(f)=\left.e^{i \Psi_{\mathrm{acc}}(f)} \tilde{h}(f)\right|_{\text {no acc }},
\end{equation}
where
\begin{equation}
\begin{aligned}
\Psi_{\mathrm{acc}}(f) & \equiv-2 \pi f X\left(z_{c}\right) \Delta T(f)^{2} \\
&=-\Psi_{N}(f) \frac{25}{768} X\left(z_{c}\right) \mathcal{M}_{z} x^{-4}
\end{aligned}
\end{equation}
corresponds to the gravitational waveform with cosmic acceleration, with $x\equiv (\pi \mathcal{M}_z f)^{2/3}$ and $\Psi_N(f) \equiv \frac{3}{128}(\pi \mathcal{M}_z f)^{-5/3}$.
The gravitational waveform without cosmic acceleration is given by
\begin{equation}
\tilde{h}(f) \big|_{\mathrm{no \ acc}}=\frac{\sqrt{3}}{2}\mathcal{A}f^{-7/6}e^{i\Psi (f)} \left[ \frac{5}{4}A_{\mathrm{pol},\alpha}(t(f)) \right] e^{-i \left( \varphi_{\mathrm{pol},\alpha}+\varphi_D \right) }, \label{waveform}
\end{equation}
with the amplitude written as
\begin{equation}
\mathcal{A}=\frac{1}{\sqrt{30}\pi^{2/3}}\frac{{\mathcal{M}_z}^{5/6}}{D_L} .  \label{amp-noangle}
\end{equation}
The polarisation amplitude $A_{\mathrm{pol},\alpha}(t)$, the polarisation phases $\varphi_{\mathrm{pol},\alpha}(t)$  ($\alpha=\mathrm{I}, \mathrm{II}$ represents the number of individual detectors \footnote{The interferometer with three arms corresponds to two individual detectors.}) and the Doppler phase $\varphi_{D}(t)$ are given in \citet{Yagi2010}. For the phase of $\Psi(f)$, we use the restricted-2PN waveform including spin-orbit coupling at the 1.5PN order \footnote{The term ``restricted'' means that we only take the leading Newtonian quadrupole contribution to the amplitude and neglect contributions from higher harmonics.} \citep{Maggiore08,Kidder93}, which is also reported by in \citet{Yagi2010}. There, we can extract the acceleration parameter $X$ and luminosity distance $D_{L}$ from the phase and amplitude of the GWs, respectively.

According to the waveform in Eq.~(6), we take the binary parameters as
\begin{equation}
\theta^{i}=\left(\ln \mathcal{M}_{z}, \ln \eta, \beta, t_{c}, \phi_{c}, \bar{\theta}_{\mathrm{S}}, \bar{\phi}_{\mathrm{S}}, \bar{\theta}_{\mathrm{L}}, \bar{\phi}_{\mathrm{L}}, D_{L}, X\right).
\end{equation}
where $\beta$ is the spin-orbit coupling parameter, and $\phi_c$ represents the coalescence phase.
$(\bar{\theta}_{\mathrm{S}}, \bar{\phi}_{\mathrm{S}})$ indicates the direction of the source in the solar barycenter frame, and $(\bar{\theta}_{\mathrm{L}}, \bar{\phi}_{\mathrm{L}})$ specifies the direction of the orbital angular momentum.

One can use Fisher analysis to estimate the measurement accuracies of the binary parameters $\theta^i$. The measurement accuracy is given by $ \Delta\theta^i  \equiv \sqrt{ \left( \Gamma^{-1}  \right)_{ii}}$~\citep{Cutler1994}, with the Fisher matrix:
\begin{equation}\label{eq5}
\Gamma_{ij}=4Re\int_{f_{\mathrm{min}}}^{f_{\mathrm{max}}}\frac{\partial_i\widetilde{h}^{*}(f)\partial_j\widetilde{h}(f)}{S_h(f)}df,
\end{equation}
where $S_h(f)$ is the analytical expression of the DECIGO noise power spectrum \citep{Kawamura2006,Kawamura2019,Yagi2011}.
The lower cut-off of frequency $f_{\mathrm{min}}=(256/5)^{-3/8} \pi^{-1} \mathcal{M}_z^{-5/8} \Delta T_{obs}^{-3/8}$ corresponds to the frequency at which coalescence begins to be observed, with $\Delta T_{obs}$ representing the observation time. And $f_\mathrm{max}$ is the higher cut-off frequency of the detector and is set equal to 100 Hz. In addition, for DECIGO, there will be eight uncorrelated interferometric signals \citep{Kawamura2011,Yagi2011}; thus, the Fisher matrix above should be multiplied by 8.

For the convenience of these calculations, we set $m_1=m_2= 1.4M_{\odot}$ and take $t_c=\phi_c=\beta=0$. For each fiducial redshift $z_c$, we randomly generate $10^4$ sets of $(\bar{\theta}_{\mathrm{S}},\bar{\phi}_{\mathrm{S}},\bar{\theta}_{\mathrm{L}},\bar{\phi}_{\mathrm{L}})$, and for each set, we calculate the uncertainty $\sqrt{\left( \Gamma^{-1} \right)_{ii}}$. Therefore, only two parameters need to be estimated: $D_L$ and $X$.
By marginalizing other parameters in the Fisher matrix, this calculation converts into a 2-Dimension submatrix of $D_L$ and $X$, with the instrumental uncertainty of $X$ being:
\begin{equation}\label{eq7}
\sigma_{X} = 8^{-1/2} \left[\left( \Gamma^{-1} \right)^{1/2}_{jj}\right].
\end{equation}
Similarly, the measurement error of luminosity distance is estimated as
\begin{equation}\label{eq7}
\sigma^{instr}_{D_L} = 8^{-1/2} \left[\left( \Gamma^{-1} \right)^{1/2}_{ii}\right].
\end{equation}
Considering the lensing uncertainty caused by the weak lensing effect $\sigma^{lens}_{D_L}=0.05zD_L$ \citep{Sathyaprakash2010}, the luminosity distance error is considered to be:
\begin{equation}\label{eq8}
\sigma_{D_L}=\sqrt{(\sigma^{instr}_{D_L})^2+(\sigma^{lens}_{D_L})^2}.
\end{equation}

\begin{figure}
\begin{center}
\includegraphics[width=0.95\linewidth]{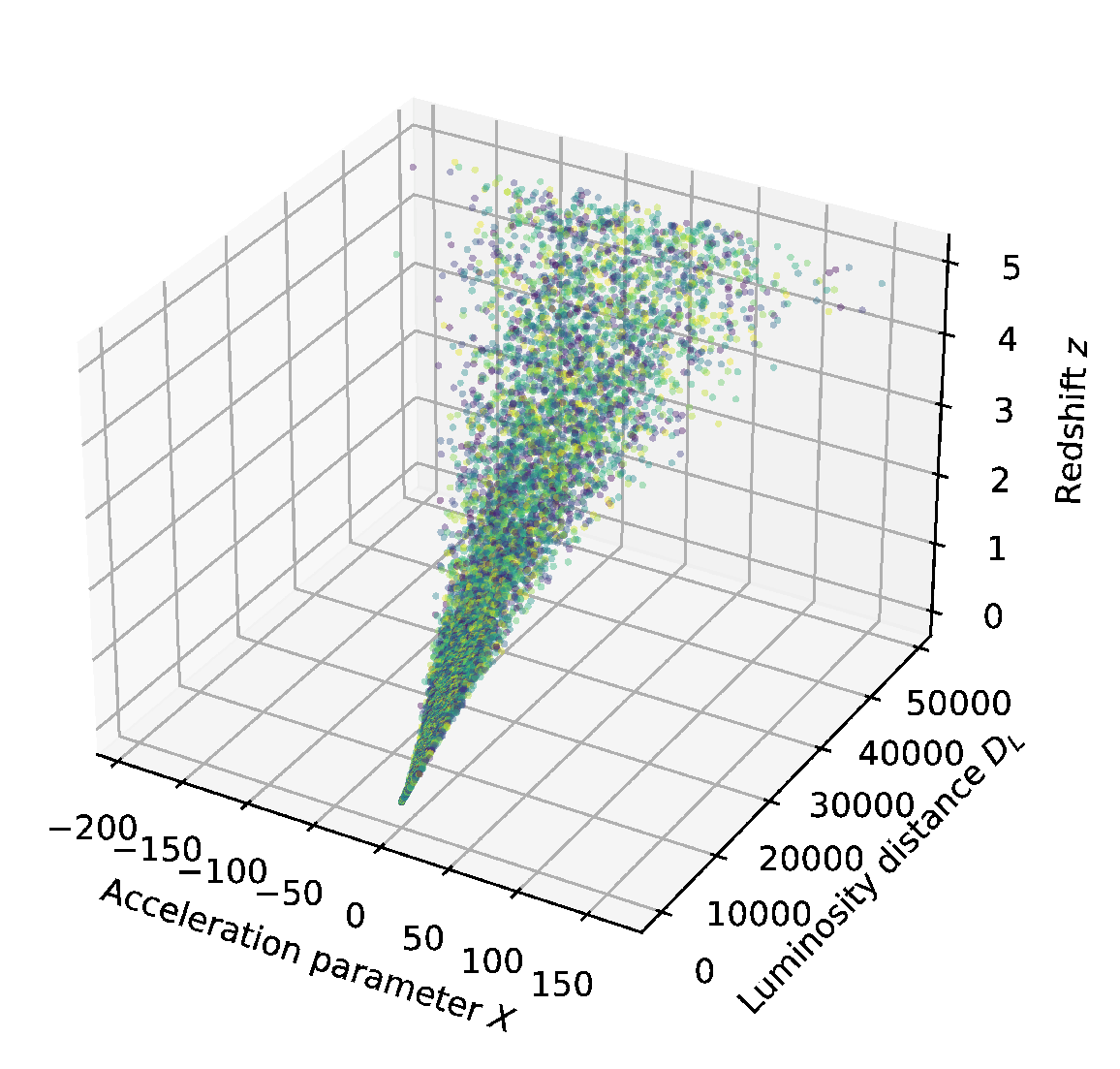}
\end{center}
\caption{3D plot of the central values of acceleration parameters and luminosity distances at different redshifts, considering 10,000 GW events detected by DECIGO.}
\end{figure}

\begin{figure}
\begin{center}
\includegraphics[width=0.95\linewidth]{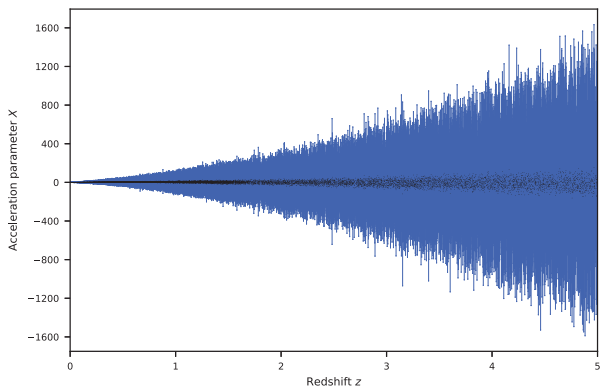}
\end{center}
\caption{Acceleration parameters (black dots) and the corresponding errors (blue bars) observed by DECIGO.}
\end{figure}

For the redshifts of GW sources, we sample them from the merger rate of double compact objects that reflects the star formation history
\citep{Dominik2013}, taking the data from the so-called ``rest frame rates'' in the cosmological scenario \footnote{http:www.syntheticuniverse.org}. In our simulation, we assume a flat $\Lambda$CDM as the proposed fiducial model with the cosmological parameters derived by {\it Planck2018} measurements \citep{Planck Collaboration2018}. The central values of the simulated 10,000 $X(z)$ and $D_L(z)$ measurements are shown in Fig.~1, and the acceleration parameters and their corresponding errors are shown in Fig.~2.

\subsection{Redshift determination from optical follow-up observations}

Our main concern here is what fraction of binary sources can we determine the redshifts in the era of DECIGO. For neutron-star (NS) binaries, the most reliable method is determining the electromagnetic counterpart of GW events directly. Meanwhile, benefit from the angular resolution of $\sim1$ arcsec$^{2}$, DECIGO is expected to uniquely identify the host galaxy of the binary \citep{Cutler2009}, the redshift of which could be determined from multi-messenger EM observations.

We now estimate the number of potential host galaxies with redshift determination for the binaries. Following the methodology proposed in \citet{Holz2005}, there would be more than $\sim 10^{11}$ galaxies potentially observable over the entire celestial sphere, with the number density of galaxies $\sim 10^{3}$ /arcmin$^{2}$ inferred from the observation of the Hubble Ultra Deep Field (HUDF) \citep{Beckwith2006, Cutler2009}.
While the future galaxy spectroscopic surveys, JDEM/WFIRST \citep{Gehrels2010, Michael2011}, plan to obtain spectroscopic redshifts for $10^{8}$ galaxies in the redshift range 0.5$< z<$2 with a precision better than 0.1$\%$ (the number of galaxies with photometric redshift measurements is expected to be larger, $\sim10^{9}$). Thus, the fraction of galaxies whose redshifts are listed in the galaxy catalogue is $\sim 10^{-3}$. DECIGO is expected to observe $10^{6}$ GW events coming from the neutron star binaries within redshift $z\sim5$ \citep{Kawamura2019,Nishizawa2012}. If the GW events randomly occur in any one of the galaxies, there would be $\sim 10^{3}$ events with redshift determination ($\sim10^{4}$ if photometric galaxies are considered). In addition, the proposed wide-field survey BigBOSS \citep{Schlegel2010} would measure $\sim$ 5 million spectroscopic redshifts per year for galaxies in the range 0.2$< z <$3.5. In the era of DECIGO, LSST may have already determined the photometric redshifts for a large fraction of host galaxies in $\sim$ 1/3 of the sky \citep{Cutler2009}. Such methodology could be easily extended to $z\sim3$, focusing on Euclid's near-infrared photometry combined with ground-based optical photometry.

For high-redshift GWs, one could turn to high-redshift tracers such as quasars or gamma-ray bursts (GRBs). The Gamow Explorer program, which has been proposed for searching for X-ray and optical-IR counterparts to high-redshift GW events \citep{White2021}, could rapidly detect the GRB (by Lobster Eye X-ray Telescope) and provide GRB redshift estimates (by Photo-$z$ Infra Red Telescope). It also enables space and ground-based observatories to follow up this GRB to determine the redshift of its host galaxy and study the afterglow in detail \footnote{Note that such strategy has been used by Swift and ground telescopes to identify high redshift GRBs, e.g., GRB 090423 at $z\sim8.23$ \citep{Tanvir2009}.}. For the follow-up observations, the deep field of HST is useful to observe galaxies that are fainter than the characteristic brightness, which not only contributes to the observations of galaxies at a very high redshift ($z\sim11$), but also provides important observations for galaxies in the redshift range of 2$< z <$5 \citep{Beckwith2006}. As a scientific successor to HST, James Webb Space Telescope (JWST) \citep{Gardner2006} can detect galaxies at much higher redshifts, and also observe faint infrared afterglows of short GRBs from BNS mergers at a distance of 150 Mpc \citep{Lu2021}, and kilonovae within a distance of $\sim$200 Mpc \citep{Bartos2016}. Other ground-based facilities would also be alerted a few years before the mergers, including Keck Observatory, Gran Telescopio Canarias (GTC), Gemini, Very Large Telescope array (VLT) \citep{Hartoog2015} and future planned 40 m facilities such as E-ELT. We remark here that DECIGO can observe GW signals several years before the coalescence, and predict coalescence time with an accuracy of $\sim0.1$ sec \citep{Kawamura2021}. Based on the precise time and sky location of GW events months in advance, simultaneous gamma-ray observations and electromagnetic follow-up observations would be more reliable and frequent. Therefore, multi-messenger astronomy will develop significantly under the DECIGO  framework \citep{Chen2021,Cao2022}.

In the above estimation, we only provide the fraction of galaxies whose redshifts are listed in the galaxy catalogue. Such a worst case with large uncertainty could be markedly improved, with dedicated follow-up observations targeted at the GW events or the host galaxies. Actually, the fraction of neutron-star binaries with redshift determination would be much larger, considering the fact that GW events are likely to occur in more massive luminous galaxies that are easier to observe. In this sense, it is reasonable to consider a conservative case during analysis, with the simulation of 10,000 GW events used to derive numerical constraints and redshift reconstruction of cosmic curvature.

\subsection{Numerical constraints on cosmic curvature $\Omega_k(z)$}

GW can provide direct measurements of the luminosity distance $D_L(z)$ and the cosmic acceleration $X(z)$, as we simulated in Section 2.1. Then, one can simply confront the distance $D_{L}(z)$ with theoretical distances $D_{L}^{X}(z)$ inferred from Eq.~(2), where curvature parameter $\Omega_{k}$ is considered and $E(z)$ can be derived from the definition equation of $X(z)$ in Eq.~(5), to provide a numerical constraint on the cosmic curvature $\Omega_k(z)$. This process could be achieved by minimizing the $\chi^{2}$ statistic
\begin{equation}
\chi^2(\Omega_{k})=\sum \limits_{i=1}^{10,000}\frac{\left[D_{L,i}^{X}(\Omega_{k}) - {D_{L,i}}\right]^{2}}
{\sigma_{D^{X}_{L},i}^{2}+\sigma_{D_{L},i}^{2}}\;,
\end{equation}
where $\sigma_{D^{X}_{L}}$ can be derived from Eq.~(2) and $\sigma_{D_{L}}$ is given in Eq.~(14). We use the Markov Chain Monte Carlo (MCMC) method \citep{Foreman2013} to obtain the best-fit value of $\Omega_{k}$ and its uncertainty, which is a model-independent estimation of the cosmic curvature.

\subsection{Individual measurements of cosmic curvature $\Omega_k(z)$}

Future space-based GW detectors, such as DECIGO, will aim to detect a large number of neutron-star binaries at much higher redshifts, which enables the reconstruction of cosmic curvature in the early universe. The cosmic curvature $\Omega_k$ can be expressed as \citep{Clarkson2007}
\begin{equation}\label{eq3}
\Omega_k=\frac{[H(z)D^{'}(z)]^2-c^2}{[H_0D(z)]^2},
\end{equation}
where the expansion rate of universe $H(z)$ can be obtained from the GW measurement of $X(z)$, the transverse comoving distance $D(z)$ is simply related to the luminosity distance $D_L(z)$ as $D(z)=D_L(z)/(1+z)$ \citep{Hogg1999}, and $D^{'}(z)=dD(z)/dz$ denotes the derivative of $D(z)$ with respect to redshift $z$. More specifically, we use the Gaussian processes (GP) method \citep{Seikel2012a} to reconstruct the first derivation of luminosity distance $D_L(z)$. This method, which assumes that the distribution of data is Gaussian, can effectively reconstruct a function and its derivatives from a given data set without parametrization \citep{Shafieloo12,Cao2019a,Qi2019a,Liu19,Wu2020,Zheng2020}. The reconstruction of $D_L'(z)$, together with the observations of $D_L(z)$ and $X(z)$ at individual redshifts, will provide different measurements of $\Omega_{k}$ through Eq.~(16).

It is necessary to give a brief introduce to the Gaussian process, which is executed by the Python package Gapp \footnote{https://github.com/carlosandrepaes/GaPP} in this work. Given a data set $\mathcal{D}$ of observations: $\mathcal{D} =\{(x_i,y_i)\}$ (in the cases investigated in this study, $x$ and $y$ are the redshift $z$ and the luminosity distance $D_L$ from the simulation, respectively). The covariance function cov$(f(x)$, $f$($\tilde{x}$)) = $k$($x$, $\tilde{x}$) is used to describe the connection between the function value at $x$ and the function value at the other point $\tilde{x}$.
Here, we consider the squared exponential covariance function
\begin{equation}
k(x,\tilde{x})=\sigma^2_{f}exp(-\frac{(x-\tilde{x})^2}{2\ell^2}),
\end{equation}
where the characteristic lengths $\ell$ and $\sigma_{f}$ represent the typical changes in $f(x)$ in x-direction and y-direction, respectively. These two hyperparameters can be trained by the observational data.

First, we reconstruct a function $f(x)$ to describe the data set, which can be expressed by the mean value $\mu(x)$ and the covariance function cov$(f(x)$, $f$($\tilde{x}$)):
\begin{equation}
f(x) \sim \mathcal{GP}\left( \mu(x), k(x,\tilde{x})
\right) \;.
\end{equation}
However, there is a difference between the real data $y$ and the reconstructed Gaussian function $f(x)$: $y_i =f(x_i) + \epsilon_i$, where Gaussian noise $\epsilon_i$ with variance $\sigma_i^2$ is assumed. Therefore, for a set of observational points $X=\left\{x_i \right\}$, the observational data can be written as
\begin{equation}
y \sim \mathcal{GP}(\mu, K(X, X)+C),
\end{equation}
by adding the variance to the covariance matrix $C$ of the data.

For observations with a limited sample size, one may need to reconstruct another function $f^*$ to describe the observational data but at some other points $X^*$, which are typically an extended point set. Simply, we obtain
\begin{equation}
f^* \sim \mathcal{GP}(\mu^*,K(X^*, X^*)),
\end{equation}
where $\mu^*$ is a prior assumed mean of $f^*$.
Combining these two Gaussian processes for $y$ and $f^*$ (Eq.~(19) and Eq.~(20)) and calculating the conditional distribution, we can reconstruct the mean and covariance of $f^*$ by
\begin{equation}
\overline{f^*}=\mu^*+K(X^*, X)[K(X, X)+C]^{-1}(y-\mu)
\end{equation}
and
\begin{equation}
{\text{cov}}(f^*)=K(X^*, X^*)-K(X^*, X)[K(X, X)+C]^{-1}K(X, X^*).
\end{equation}
And the variance of $f^*$ can be simply obtained by diagonalizing $\text{cov}({f^*})$. Thus, we can expand the observational data set.

The derivative of the function $f^*$ can also be calculated by giving the Gaussian process for $y$ and ${f^*}'$. Similarly, we have
\begin{equation}
\overline{{f^*}'} = \text{$\mu^*$}' + K'( X^*, X)
\left[K( X, X) + C\right]^{-1} ({y}-\text{$\mu$})
\end{equation}
and
\begin{equation}
\begin{split}
\text{cov}({f^*}') = - K'(X^*,X)\left[K(X,X) + C \right]^{-1}  K'(X,X^*)\\
+ K''(X^*,X^*),
\end{split}
\end{equation}
where
\begin{equation}
[K'(X,X^*)]_{ij} = \text{cov}\left(f_i,\frac{\partial f^*_j}{\partial x^*_j} \right)=\frac{\partial k(x_i,x^*_j)}{\partial x^*_j}
\end{equation}
is the covariance between the function $f^*$ and its derivative,
and
\begin{equation}
[K''(X^*,X^*)]_{ij} = \text{cov}\left(\frac{\partial f^*_i}{\partial x^*_i}, \frac{\partial
  f^*_j}{\partial x^*_j} \right)=\frac{\partial^2k(x^*_i,x^*_j)}{\partial
    x^*_i\,\partial x^*_j} \;.
\end{equation}
is the covariance between its derivatives. Then both function $f^*$ and its derivative ${f^*}'$ can be reconstructed from the observational data $\mathcal{D}$.

\section{Results and discussion}

We first combine the $D_{L}(z)$ and $X(z)$ to provide a numerical constraint on $\Omega_k$ by calculating the $\chi^{2}$ statistics in Eq.~(15). The result from 10,000 simulated GW events detected by DECIGO is
\begin{equation}
\Omega_k= -0.05 \pm 0.12.
\end{equation}
This result is consistent with the fiducial value of $\Omega_k = 0$ within a 1$\sigma$ confidence level.
Compared to the other model-independent tests involving the $H(z)$ and distances from popular cosmological probes, our result is more precise than that of the Pantheon SNe Ia plus $H(z)$ (CC) as $\Omega_k= 0.63 \pm 0.34$ \citep{Wang2020a}, and the result of UV+Xray quasars plus $H(z)$ (CC) as $\Omega_k=-0.92 \pm 0.43$ \citep{Wei2020}.

\begin{figure}
\begin{center}
\includegraphics[width=0.95\linewidth]{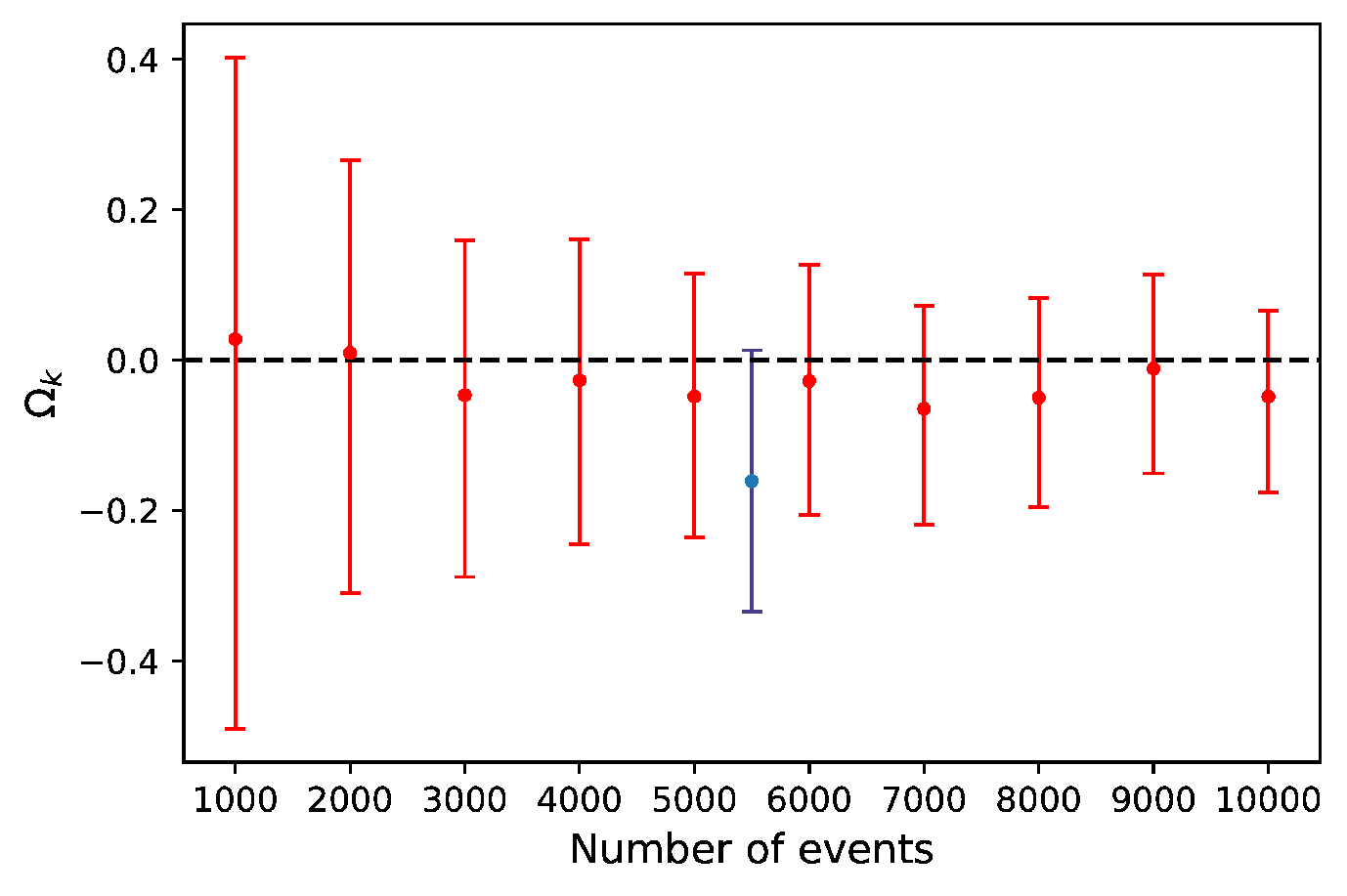}
\end{center}
\caption{Best-fit $\Omega_{k}$ and $1\sigma$ confidence level as a function of the number of GW events. The blue diamond represents the model-independent constraint from Pantheon SNe Ia data combined with the cosmic chronometers $H(z)$. The black dashed line is the fiducial value.}
\label{f3}
\end{figure}

In order to demonstrate the precision of the curvature parameter assessment with a certain number of GW events,
we show the best-fit $\Omega_{k}$ and $1\sigma$ confidence level as a function of the number of GW events in Fig.~\ref{f3} and Table~\ref{table1}. The model-independent test of $\Omega_{k}$ from Pantheon SNe Ia (blue diamond), which will be introduced in the later analysis, is also plotted for comparison. As one may see, the precision of the determined $\Omega_{k}$ from 6000 GW events ($\Delta \Omega_k=0.165$) is more competitive than that of Pantheon SNe Ia sample.

\begin{table}
\centering \caption{Summary of model-independent curvature determinations from different number of GW events.}
\begin{tabular}{cc|cc}
\hline
\hline
 $N$ &  $\Omega_{k}$  &  $N$  &  $\Omega_{k}$ \\
\hline
1000   &   $0.03^{+0.52}_{-0.37}$   &   6000   &   $-0.03\pm0.17$   \\
2000   &   $0.01\pm0.29$   &   7000   &   $-0.07\pm0.15$   \\
3000   &   $-0.05\pm0.22$   &   8000   &   $-0.05\pm0.14$   \\
4000   &   $-0.03\pm0.20$   &   9000   &   $-0.01\pm0.13$   \\
5000   &   $-0.05\pm0.18$   &   10000   &   $-0.05\pm0.12$   \\
\hline
\end{tabular}
\label{table1}
\medskip \\
\end{table}

The individual measurements of 10,000 $\Omega_{k}(z)$ obtained from Eq.~(16) in Section 2.4 are shown in Fig.~4. To estimate how much improvement of the combing measurement, in Fig.~4 we also summarize the multiple $\Omega_{k}$ within the redshift bin of $\Delta z=0.1$ through the inverse variance weighting, which allows a direct check of its predicted constancy with the given redshift. We find that with the increase of redshift, the derived $\Omega_{k}$ remains within the error bar (68.3\% confidence level [C. L.]) of the flat case, which underlies the assumption of our GW data simulations.
It is worth noticing that the uncertainty of $\Omega_{k}$, which is much higher in the low redshift range, also fluctuates at high redshifts ($3<z<5$). Such a tendency could be explained by the term $[(D'(z)/D(z))^{2} E(z) \sigma_{E(z)}]$ in the $\Omega_{k}(z)$ error equation derived from Eq.~(16), which dominates the uncertainty of $\Omega_{k}$ measurements. In this term, the function of $(D'(z)/D(z))^{2}$ with large values at $z\sim 0$ tends to decrease with increasing redshift, which generates a relatively small $\Omega_{k}$ uncertainty at higher redshifts. However, the function of $[E(z) \sigma_{E(z)}]$ exhibits an opposite tendency, generating a fluctuation of $\Omega_{k}$ uncertainty at higher redshifts when combined with the function of $(D'(z)/D(z))^{2}$.

\begin{figure}
\begin{center}
\includegraphics[width=0.95\linewidth]{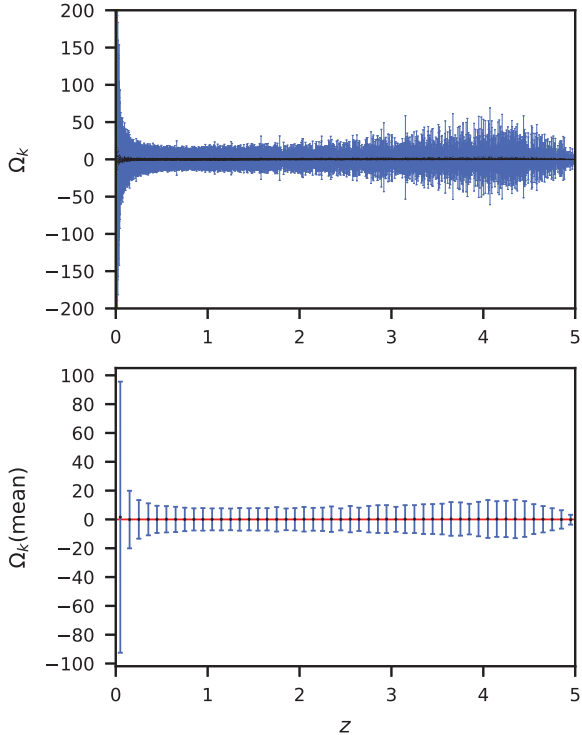}
\end{center}
\caption{Individual measurements of cosmic curvature (upper panel) and their redshift-binned counterparts (lower panel) from the standard sirens observed by DECIGO.}
\end{figure}

For comparison, we also provide model-independent measurements of $\Omega_{k}$ from electromagnetic observations. The 1048 luminosity distance measurements from SNe Ia Pantheon, as well as 41 observational Hubble parameter data (OHD) from cosmic chronometer (CC) and baryon acoustic oscillation (BAO) measurements \citep{Gaztaaga2009,Blake2012,Busca2013,Samushia2013,Xu2013,Font-Ribera2014,Delubac2015}, are used to test the cosmic curvature. Let us note that in order to achieve a better redshift match with the supernova data, we use GP to obtain
the reconstruction of a smooth $H(z)$ function, with the generation of 200 reconstructed $H(z)$ data well matched the redshifts of Pantheon SNe Ia sample.

According to the BEAMS with Bias Corrections (BBC) method \citep{Kessler2017}, the observed distance modulus of SNe can be simply given by 
the apparent ($m_B$) and absolute \textit{B}-band magnitude ($M_B$) as $\mu=m_B - M_B$, with the nuisance parameters in the Tripp formula \citep{Tripp1998} retrieved. The theoretical distance modulus $m_{th}$ can then be obtained by
\begin{equation}
m_{th} =5\log\frac{D_L^{th}}{\rm Mpc}+25 + M_B,
\end{equation}
where $D_L^{th}$ is given by Eq.~(2) involving the reconstructed $H(z)$. Then, the cosmic curvature $\Omega_k$ and the parameter $M_B$ can be constrained by minimizing the $\chi^{2}$ statistic
\begin{equation}
\chi_{SNe}^2=\sum \limits_{i=1}{\frac{[m_i^{obs}-
m_i^{th}]^2}{\sigma_{SNe}^2+\sigma_{m_{th}}^2}},
\end{equation}
where $\sigma_{SNe}$ accounts for the error in SNe Ia observations, propagated from the covariance matrix in \citet{Scolnic2018}. The marginalized probability distribution of each parameter and the marginalized 2-D confidence contours are shown in Fig.~5. The best-fit cosmic curvature and the absolute \textit{B}-band magnitude with $1\sigma$ are $\Omega_k=-0.16\pm0.17$ and $M_B=-19.32\pm0.01$. This result favors a zero cosmic curvature at 68.3$\%$ confidence level, which shows no evidence for the deviation from the flat universe at the current observational data level. Meanwhile, the uncertainty of $\Omega_k$ from the GW method ($\Delta \Omega_k \sim 0.12$) is $\sim 30\%$ smaller than that from the SNe Ia method. Moreover, the SNe Ia method is strongly dependent on the choice of the reconstruction methods of $H(z)$ function, as was pointed out in the recent works of \citet{Wang2020a}. The determined cosmic curvature $\Omega_k$ is strongly degenerated with the absolute magnitude $M_B$ of SNe Ia, similar to the results obtained by examining the cosmic opacity with gravitational waves and Type Ia Supernova \citep{Qi2019c}.

\begin{figure}
\begin{center}
\includegraphics[width=0.95\linewidth]{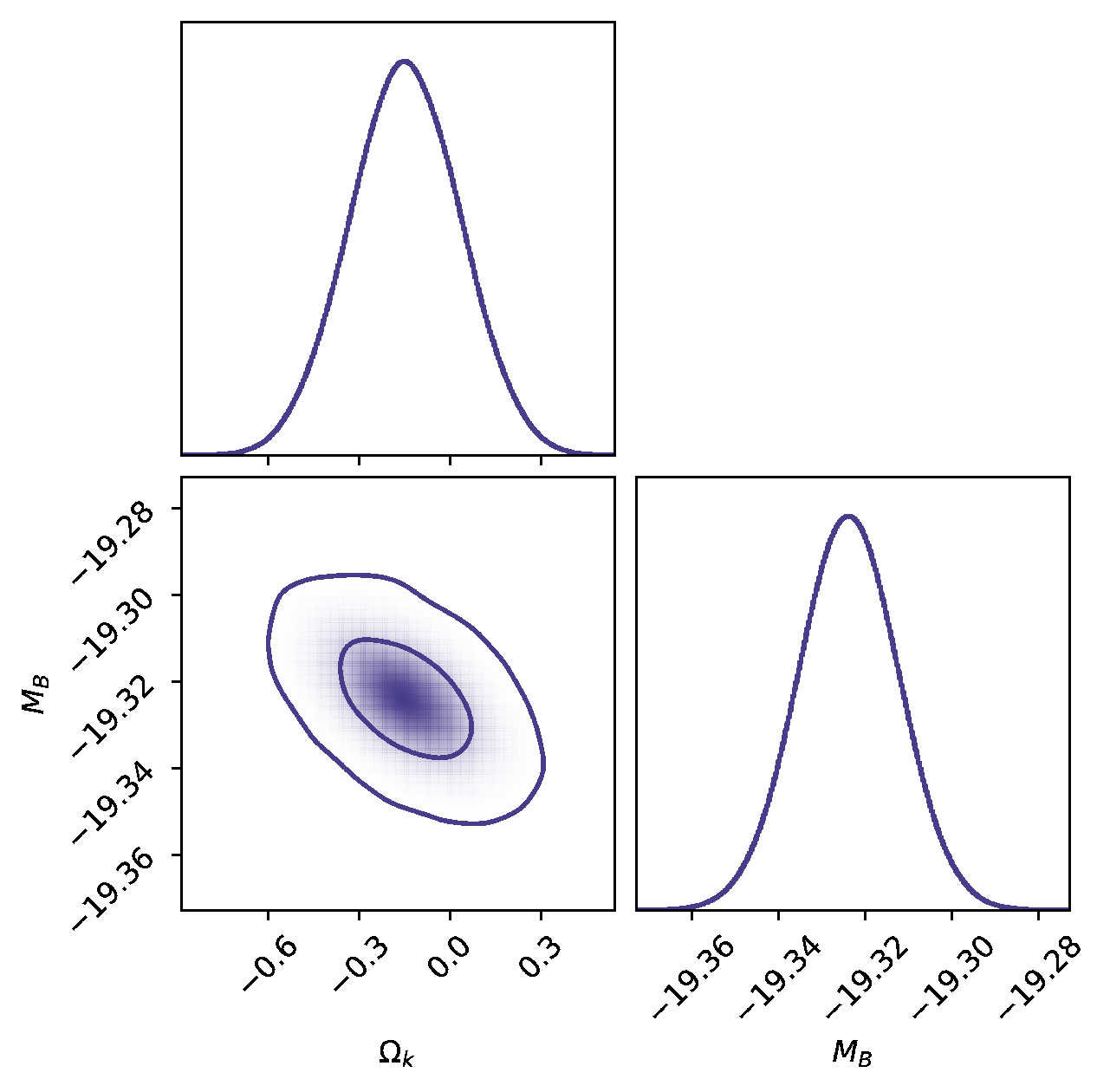}
\end{center}
\caption{Marginalized constraints on $M_B$ and $\Omega_k$ from the Pantheon SNe Ia and OHD data.}
\end{figure}

Then, we transform the distance modulus $\mu$ to $D_L$ through
\begin{equation}
D_{L}(z)=10^{\mu(z)/5-5} (\textmd{Mpc}),
\end{equation}
where the absolute magnitude of SNe Ia is set at the best-fit value $M_B=-19.32$ from the results shown in Fig.~5. Following the same procedure as the GW observations, the cosmic curvature is measured at 41 different redshifts, as shown in Fig.~6. Compared with the GW method, the redshift of $\Omega_k$ is only up to $ z \sim$ 2.3, limited by the redshift coverage of SNe Ia (0.01$<z<$2.26) and OHD (0.07$<z<$2.36). Meanwhile, the $\Omega_k$ measurement based on our GW method is more precise than those from the current EM observations, which indicates another advantage of our methodology in testing the spatial curvature in the GW domain.

The performance of the current traditional EM method is restricted by the big gap between the amount, as well as the redshift coverage of the observational $H(z)$ data and $D_L(z)$ data. On the one hand, the Hubble diagram of type Ia supernovae (SNe Ia) contains only the order of $\sim 10^3$ SNe Ia. Such situation will be greatly improved in the era of the Large Synoptic Survey Telescope (LSST), which could discover an unprecedented number of SNe Ia $\sim 10^6$, with a large fraction ($\sim10\%$) expected to be turned into useful distance indicators \citep{Lochner2021}. On the other hand, future observations of redshift drift \citep{Sandage1962}, which is also known as the Sandage-Loeb test, provide an important method to derive precise measurements of $H(z)$ at different redshifts. Specifically, by observing the redshift drift in the optical and radio bands, the European Extremely Large Telescope (E-ELT) and the Square Kilometre Array (SKA) will offer the $H(z)$ measurements in the redshift ranges of $2<z<5$ and $0<z<0.3$, respectively \citep{Liske2008,Quercellini2012,Martins2016}. In addition, strongly lensed SNe Ia, which will also be discovered in larger numbers by LSST, enables a more precise model-independent probe of cosmological parameters based on the distance sum rule \citep{Cao18,Ma19b}. Specially, based on the simulated sample of 200 lensed SNe Ia with time-delay measurements \citep{Qi2022}, model-independent constraints of the Hubble constant $H_0$ and cosmic curvature parameter $\Omega_k$ would be achieved with high precision ($\Delta H_0$ = 0.33 $km s^{-1} Mpc^{-1}$ and $\Delta\Omega_k$ = 0.05). Therefore, one might be optimistic about achieving much higher precision of improved EM observations in the next decades. In that case, the precision of $\Omega_k$ measurements in the EM domain would be much higher. Whereas, considering the the difficulty of deriving multiple measurements ($D_L(z)$ and $H(z)$) exactly at the same redshift, the prospects for
constraining the cosmic curvature in the GW domain could be promising, based on the combination of two different observables for the same objects at high redshifts.

\begin{figure}
\begin{center}
\includegraphics[width=0.95\linewidth]{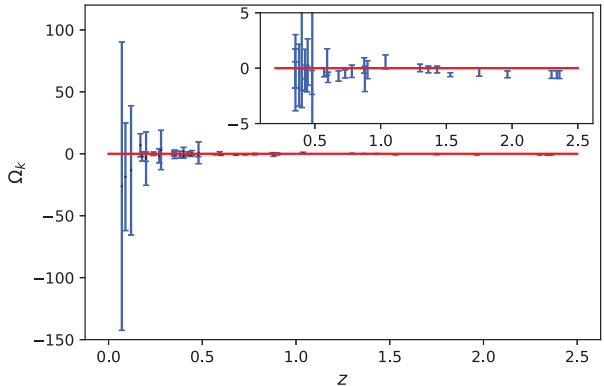}
\end{center}
\caption{Individual measurements of cosmic curvature from the current Pantheon SN and OHD data.}
\end{figure}

\section{Conclusions}

As a standard siren, gravitational wave (GW) from binary neutron star merger provides a direct way to measure the luminosity distance ($D_L$) without the need of cosmological distance ladder. In addition, the accelerating expansion of the universe may cause an additional phase shift in the gravitational waveform, which allows us to measure the acceleration parameter. Thus, GW measurement provides an important opportunity to determine the curvature parameter $\Omega_k$ in the GW domain based on the combination of two different observables for the same objects at high redshifts.

In this paper, we investigate how such an idea could be implemented with future generation of space-based DECi-hertz Interferometer Gravitational-wave Observatory (DECIGO) in the framework of two model-independent methods. Our results show that DECIGO could provide a reliable and stringent constraint on the cosmic curvature at a precision of $\Delta\Omega_k$=0.12, which is comparable to the latest model-independent estimations using different EM probes. Furthermore, we use the GP method to reconstruct the first derivative of $D_L$ and obtain individual 10,000 measurements of $\Omega_k$ at different redshifts ($z\sim 5$). Compared to the traditional model-independent estimations of the spatial curvature using other EM observations, GW sirens have several benefits as follows:

\begin{itemize}
\item Cosmological-model independent: GW standard siren measurements could provide independent measurements of luminosity distance and acceleration parameter without the cosmological distance ladder. And the matched filtering analysis in GW method does not require an assumption of any fiducial cosmological model. In addition, the $\Omega_k$ from EM method is strongly degenerate with other parameters such as the absolute magnitude, while that from GW standard sirens has the advantage of no nuisance parameters.
\item High redshift: GW detectors can observe a large number of events at high redshifts, which allows us to probe the cosmic curvature at high redshifts. However, the current OHD obtained by radial baryon acoustic oscillations (BAO) and cosmic chronometer (CC) still provide no high redshift data.
\item Well redshift matched: Both the luminosity distance and the acceleration parameter can be determined from each GW event, which means that they are already well matched in redshift and can be used directly for the statistical constraint on $\Omega_k$ (as Eq.(15)). While the traditional EM method, due to the large gap between the amount of observational $H(z)$ data and observational $D_L(z)$ data, a parametric or non-parametric method must be used to reconstruct $H(z)$ before, restricting the constraint ability of SNe Ia. In addition, the redshift match between two sets of data also leads to errors.
\end{itemize}

However, there are still some issues that should be emphasized here:

\begin{itemize}
\item The redshifts of the GW sources are necessary ingredients to perform model-independent constraints on the cosmic curvature with DECIGO. Considering the angular resolution of DECIGO ($\sim1$ arcsec$^{2}$), which can uniquely identify the host galaxy of the binary, we could take a widely used method that through the optical (or infrared) identification of the host galaxy of the GW event. For high-redshift GWs, one could turn to high-redshift tracers such as quasars or gamma-ray bursts (GRBs), along with follow-up observations to determine their redshifts, considering the significant development of multi-messenger astronomy in the framework of DECIGO. In addition, several methods have been proposed in the literatures to address this issue, such as the "galaxy voting" method (redshift distribution for host galaxies) \citep{MacLeod2008,Trott2021}, the redshift distribution for coalescing sources \citep{Ding2019}, neutron star mass distribution \citep{Taylor2012a,Taylor2012b}, cross-correlation of gravitational wave standard sirens and galaxies \citep{Oguri2016}, and the tidal deformation of neutron stars \citep{Messenger2012,Messenger2014,Wang2020}.
\item Different from current observational $H(z)$ data, the measurement of acceleration parameter $X$ is based on time measurement in observer coordinate (which is similar to the measurement of redshift drift), while the OHD relies on the time measurement in the university coordinate. And it appears in the 4th-PN order correction in GW waveform, which requires high-precision detection of GW signal especially at lower frequencies \citep{Seto2001, Nishizawa2012}. Fortunately, as one of its major objectives, DECIGO is competent for direct measurement of the universe acceleration. In the future, DECIGO will detect a large number of neutron-star binaries in inspiraling phases, which will provide an unprecedented opportunity for high-precision detections of cosmic acceleration and will open up a window for gravitational-wave cosmology.
\end{itemize}

Summarily, the GW observations provide a powerful and novel method to estimate the spatial curvature in different cosmological-model-independent ways. This strengthens the probative power of our method, especially in the framework of DECIGO, to inspire other new observing programs and theoretical works in the near future.

\section*{Acknowledgments}

This work was supported by the National Natural Science Foundation of China under Grants Nos. 12021003, 11690023, and 11920101003; the National Key R\&D Program of China No. 2017YFA0402600; the Strategic Priority Research Program of the Chinese Academy of Sciences, Grant No. XDB23000000; and the Interdiscipline Research Funds of Beijing Normal University.

\end{document}